\documentclass[11pt,twoside,a4paper,cmspaper,final,collab]{cms-tdr}
\def\svnVersion{5ad6014}\def\svnDate{2026/08/07}\def\cmsCernNoTag{CERN-EP-2026-051}\def\cmsCernDate{\today}\def\cmsMessage{Submitted to Physical Review Letters}
\begin{document}\cmsNoteHeader{EXO-24-012}

\newcommand{\ptvis}{\ensuremath{\pt^\text{vis}}\xspace}
\newcommand{\ptvistauh}{\ensuremath{\ptvis(\tauh)}\xspace}
\newcommand{\ptvismutauh}{\ensuremath{\ptvis(\PGm,\tauh)}\xspace}
\newcommand{\pmu}{\ensuremath{p(\PGm)}\xspace}
\newcommand{\yvismutauh}{\ensuremath\abs{y_{\text{vis}}(\mu,\tauh)}\xspace}
\newcommand{\mvis}{\ensuremath{m_{\text{vis}}}\xspace}
\newcommand{\mvismutauh}{\ensuremath{\mvis(\PGm,\tauh)}\xspace}
\newcommand{\taumu}{\ensuremath{\PGt_{\PGm}}\xspace}
\newcommand{\zetavis}{\ensuremath{\zeta^\text{vis}}\xspace}
\newcommand{\zetaptmiss}{\ensuremath{\zeta^{\ptmiss}}\xspace}
\newcommand{\taunet}{\ensuremath{\textsc{TauNet}}\xspace}
\newcommand{\HLVmu}{\ensuremath{\textsc{HLV}_{\PGm}}\xspace}
\newcommand{\Zgstar}{\ensuremath{\PZ/\PGgst}\xspace}
\newcommand{\hpm}{{\HepParticle{\PSh}{}{\pm}}\Xspace}
\newcommand{\hmp}{{\HepParticle{\PSh}{}{\mp}}\Xspace}

\ifthenelse{\boolean{cms@external}}{
    \def\appendixname{End Matter}
    \providecommand{\cmsTable}[1]{#1}
}{
    \providecommand{\cmsTable}[1]{\resizebox{\textwidth}{!}{#1}}
}

\cmsNoteHeader{EXO-24-012}

\title{Search for low-mass resonances decaying to \texorpdfstring{$\PGt\PGt$}{ditau} in proton-proton collisions at \texorpdfstring{$\sqrt{s} = 13.6\TeV$}{sqrt(s) = 13.6 TeV}}

\date{\today}

\abstract{
An inclusive search is presented for spin-zero bosons decaying to $\PGt\PGt$ in a previously unexplored mass range between 20 and 60\GeV using proton-proton collision data at $\sqrt{s}=13.6\TeV$, corresponding to an integrated luminosity of 61.9\fbinv recorded by CMS in 2022--2023. The search is performed using a high-rate trigger stream in combination with a novel low-momentum hadronic tau reconstruction algorithm. No significant excess above the standard model background is observed. Upper limits on the product of the spin-zero resonance production cross section and branching fraction to $\PGt\PGt$ are set at 95\% confidence level between 40 and 400\unit{pb}.
}

\hypersetup{
pdfauthor={CMS Collaboration},
pdftitle={Search for low mass resonances decaying to tau tau in proton-proton collisions at sqrt(s) = 13.6 TeV},
pdfsubject={CMS},
pdfkeywords={CMS, scouting, taus}}

\maketitle 

The discovery of the Higgs boson~\cite{20121,201230,chatrchyan_2013_observation} completed the standard model (SM) of particle physics, providing insights into electroweak symmetry breaking. Despite the success of the SM, numerous phenomena in particle physics remain unexplained~\cite{Fukuda_1998, PhysRevLett.126.141801, Aaij:2021vac}. 
Many theories that extend the SM in order to address these phenomena predict the existence of additional spin-zero bosons (\PGf). These particles may be less massive than the Higgs boson and have enhanced interactions with third-generation SM particles~\cite{Fayet:1974pd,Fayet:1977yc}, including decays to tau leptons~\cite{model_lep}. Consequently, searches for new low-mass resonances decaying to the $\PGt\PGt$ final states directly probe beyond-the-SM (BSM) physics at the LHC~\cite{Curtin:2013fra}.

This Letter presents a novel inclusive search for a low-mass spin-zero \PGf boson with a mass between 20 and 60\GeV produced via gluon fusion and decaying to a pair of tau leptons. Data were collected by the CMS detector from proton-proton ($\Pp\Pp$) collisions with a center-of-mass energy of 13.6\TeV at the CERN LHC in 2022--2023, corresponding to an integrated luminosity of 61.9\fbinv~\cite{LUM-22-001, CMS-DP-2024-068}. The final state used in this search consists of one of the tau leptons decaying to a muon and neutrinos (\taumu) and the other one decaying hadronically (\tauh). The analysis exploits the CMS scouting data set~\cite{CMS:2024zhe}, which retains only the event data reconstructed by the high-level trigger (HLT). The reduced event size enables higher event rates than standard triggers~\cite{CMS:2024aqx}.

Searches for the inclusive production of spin-zero particles decaying to $\PGt\PGt$ have been conducted by the CMS and ATLAS Collaborations~\cite{ATLAS:2012ube,ATLAS:2014vhc,ATLAS:2016ivh,ATLAS:2017eiz,ATLAS:2020zms,ATLAS:2024rzd,CMS:2011lzj,CMS:2012bkm,CMS:2014ccx,CMS:2015qnd,CMS:2018rmh,higg21001}. However, invariant masses below 60\GeV have never been explored at a hadron collider in the case where at least one tau lepton decays hadronically. Spin-zero particles interact with fermions through chirality violating Yukawa couplings that are proportional to the fermion mass. This renders their coupling to the initial-state particles at $\Pep\Pem$ colliders, such as the CERN LEP, negligible. For spin-zero resonance masses below 60\GeV, the tau leptons have relatively low transverse momentum (\pt) and there is substantial background from SM events composed primarily of jets produced through the strong interaction, referred to as quantum chromodynamic (QCD) multijet events.
The magnitude of this background necessitates high thresholds at the trigger level to control the data rate. Consequently, the traditional data acquisition methods have very low efficiency for low-mass $\PGt\PGt$ events. Data scouting improves the efficiency for recording events containing a \PGf of mass 20\GeV by a factor of four. Tabulated results are provided in the HEPData record for this analysis~\cite{hepdata}.

The CMS apparatus~\cite{CMS:2008xjf,CMS:2023gfb} is a multipurpose, nearly hermetic detector, designed to trigger on~\cite{CMS:2020cmk,CMS:2016ngn} and identify electrons, muons, photons, and (charged and neutral) hadrons~\cite{CMS:2020uim,CMS:2018rym,CMS:2014pgm}. A global ``particle-flow'' (PF) algorithm~\cite{CMS:2017yfk} aims to reconstruct all individual particles in an event, combining information provided by the all-silicon inner tracker, crystal-electromagnetic calorimeter (ECAL), and brass-and-scintillator hadron calorimeter (HCAL), operating inside a 3.8\unit{T} superconducting solenoid, with data from the gas-ionization muon detectors interleaved with layers of the steel return yoke outside the solenoid. The reconstructed particles (PF candidates) are used to build \PGt leptons, jets, and the missing transverse momentum vector (\ptvecmiss), computed as the negative vector sum of the transverse momenta of all the PF candidates in an event~\cite{CMS:2018jrd,CMS:2016lmd,CMS:2019ctu}. The magnitude of \ptvecmiss is denoted as \ptmiss. 

While the LHC provides collisions at 40\unit{MHz}, the CMS detector and data acquisition systems do not have the bandwidth to record events at this rate. Instead, events of interest are selected using a two-tiered trigger system. The first level (L1), composed of custom hardware processors, uses information from the calorimeters and muon detectors to select events at a rate of around 100\unit{kHz} within a fixed latency of about 4\mus~\cite{CMS:2020cmk}. The second level (HLT)~\cite{CMS:2024aqx} consists of a farm of processors running a version of the PF algorithm optimized for fast processing, and reduces the event rate to a few\unit{kHz} in the standard stream and 25\unit{kHz} in the data scouting stream before data storage~\cite{CMS:2024zhe}.

Scouting data used in this search are collected with trigger thresholds that are significantly lower than standard triggers, increasing sensitivity to BSM physics. 
The CMS scouting data stream is seeded by a set of L1 triggers targeting dimuons, electromagnetic energy deposits, and hadronic activity. To increase the HLT rate to around 25\unit{kHz}, the scouting data stream implements a faster particle reconstruction algorithm~\cite{CMS:2024zhe}. This search utilizes a logical OR of triggers selecting electromagnetic and hadronic activity, as there were no dedicated single low-\pt lepton or \tauh L1 triggers present in scouting data taken in 2022--2023. Data used in this analysis were seeded by the following L1 triggers: an energy deposit in the ECAL of at least 30\GeV, a scalar \pt sum of hadronic jets (\HT) greater than 360\GeV, or a deposit in the HCAL with $\pt>180\GeV$.

The signal sample $\Pg\Pg\to\PGf\to\PGt\PGt$ is generated using \MGvATNLO (v2.9.13)~\cite{aMCatNLO} at leading-order (LO) with up to two extra jets using the MLM matching procedure~\cite{mlmmatch}. Signal samples are produced with $m_{\PGf}$ values from 20 to 60\GeV in 10\GeV steps, allowing intermediate mass point characteristics such as shape and acceptance to be obtained by interpolation. Background samples are utilized to train neural networks that identify \taumu and \tauh candidates, validate the production cross sections of irreducible resonant backgrounds, and determine selections on data to maximize signal significance. Simulated SM sources of both genuine and misidentified $\taumu\tauh$ events are $\Zgstar$+jets, $\PW$+jets, top quark pairs (\ttbar), dibosons ($\PW\PW$, $\PZ\PZ$, $\PW\PZ$), and QCD production. The $\Zgstar\to\PGt\PGt$, $\Zgstar\to\PGm\PGm$+jets, and \ttbar processes are simulated at next-to-LO (NLO) with \POWHEG~\cite{Nason:2004rx, Frixione:2007vw, Alioli:2010xd, Alioli:2008gx, powheg_ttbar}. The QCD and diboson samples are generated at LO with \PYTHIA 8.212~\cite{pythia}.

All samples were interfaced with \PYTHIA with CP5 as the underlying event tune~\cite{cp5} using the NNPDF3.1 next-to-NLO parton distribution function set~\cite{Ball:2014uwa}. The interaction of particles with the detector is simulated using the \GEANTfour toolkit~\cite{Agostinelli:2002hh}. The effects of particles from additional $\Pp\Pp$ interactions in the same or adjacent bunch crossings (pileup) are incorporated by overlaying minimum bias events, simulated with \PYTHIA, on the hard scattering interactions, with the multiplicity distribution of these additional interactions matching that observed in data.

Standard \tauh reconstruction at CMS employs the hadrons-plus-strips (HPS)~\cite{CMS:2018jrd} algorithm and is tuned to reconstruct \tauh candidates if the magnitude of the transverse component of the vector sum of non-neutrino decay particles (\ptvis) is greater than 18\GeV. To achieve sensitivity in the phase space of interest, we developed a novel \tauh reconstruction algorithm (scouting HPS) with enhanced sensitivity for \ptvis as low as 5\GeV. Unlike standard CMS data formats, those from data scouting only contain partially reconstructed PF candidates~\cite{CMS:2017yfk}. These reconstructed particles are assembled into composite objects and checked to determine if they meet the composition and kinematic characteristics consistent with the \tauh decay. The standard HPS reconstruction algorithm limits the opening angle of the \tauh to $\Delta R<0.1$~($\Delta R=\sqrt{\smash[b]{(\Delta\eta)^{2}+(\Delta\phi)^2}}$ where $\eta$ is the pseudorapitidy and $\phi$ is the azimuthal angle). To capture \tauh candidates with low \ptvis, we implement a dynamic maximum opening angle. At \ptvistauh of 5\GeV, the $\Delta R$ of the cone is allowed to be as large as 0.4. This maximum value decreases smoothly to 0.1 for \ptvistauh above 18\GeV. We consider four types of \tauh decays: one-prong, three-prong, one-prong-plus-one-strip, and one-prong-plus-two-strips, where prongs and strips refer to charged hadrons and \PGpz respectively. Once \tauh candidates are identified, we use a ``deep sets'' model~\cite{Zaheer:2017wmg,Komiske:2018cqr} (\taunet) to generate a discriminant based on their constituents and surrounding PF candidates. We select the \taumu candidate as the highest \pt PF muon~\cite{CMS:2017yfk} in each event and then use a neural network isolation discriminant (\HLVmu) to identify muons with low neighboring activity. Both \taunet and \HLVmu are described in detail in \appendixname~\ref{App:ScoutngHPSRecoDetails}. 

Events that pass any of the trigger selection criteria are considered if they contain a PF muon candidate with \pt greater than 3\GeV. These events are subsequently processed by scouting HPS to create scouting \tauh candidates. Events are then selected based on the following variables: \ptvistauh, $\pt(\PGm)$, magnitude of the vector sum of $\ptvec^{\text{vis}}(\tauh)$ and $\ptvec(\PGm)$ referred to as \ptvismutauh, \taunet, \HLVmu, and linear combinations of the $\taumu\tauh$ system bisector projection variables \zetavis and \zetaptmiss, defined in~\cite{CDFditau}. To reject backgrounds with muons from \PW boson and heavy flavor decays, we use the transverse mass of the system formed by the reconstructed four-momenta \pmu and (\ptmiss, \ptvecmiss), defined as $\mT=\sqrt{\smash[b]{m^2(\pmu,(\ptmiss,\ptvecmiss)) + \pt^2(\pmu,(\ptmiss,\ptvecmiss))}}$~\cite{ParticleDataGroup:2024cfk}. All candidates must lie within $\abs{\eta}<2.4$ and only events with oppositely charged \tauh and \taumu candidates are considered.

We search for resonances as localized excesses in the mass distributions derived from the visible components of the $\taumu\tauh$ decay products (\mvismutauh). Two selection regions (Table~\ref{table:cuts_by_region}) are used in the analysis: one to observe the \PZ boson, and one to perform a search for the \PGf boson. The observation of the \PZ boson at the expected rate confirms our ability to reconstruct \tauh decays and provides a control sample for determining uncertainties in the \PGf search region. 

\begin{table*}[htb!]
\centering
\topcaption{Summary of selection criteria for the \PGf boson and \PZ boson regions.}
\cmsTable{
\begin{scotch}{lccccccc}
    Region & \ptvistauh & $\pt(\PGm)$ & \ptvismutauh & \taunet & \HLVmu & \mT & $\zetaptmiss-0.85\zetavis$ \\
      & [{\GeVns}] & [{\GeVns}] & [{\GeVns}] & & & [{\GeVns}] & [{\GeVns}]\\
    \hline
    \PGf & $>$7  & $>$5 & $>$20 & $>$0.88 & $>$0.95 & $<$100 & \NA \\
    \PZ & $>$20 & $>$20 & $>$0 & $>$0.9 & $>$0.9 & $<$100 & $>\!-25$ \\
  \end{scotch}
  }
  \label{table:cuts_by_region}
\end{table*}

\begin{figure}[htpb]
    \centering
    \includegraphics[width=0.45\textwidth]{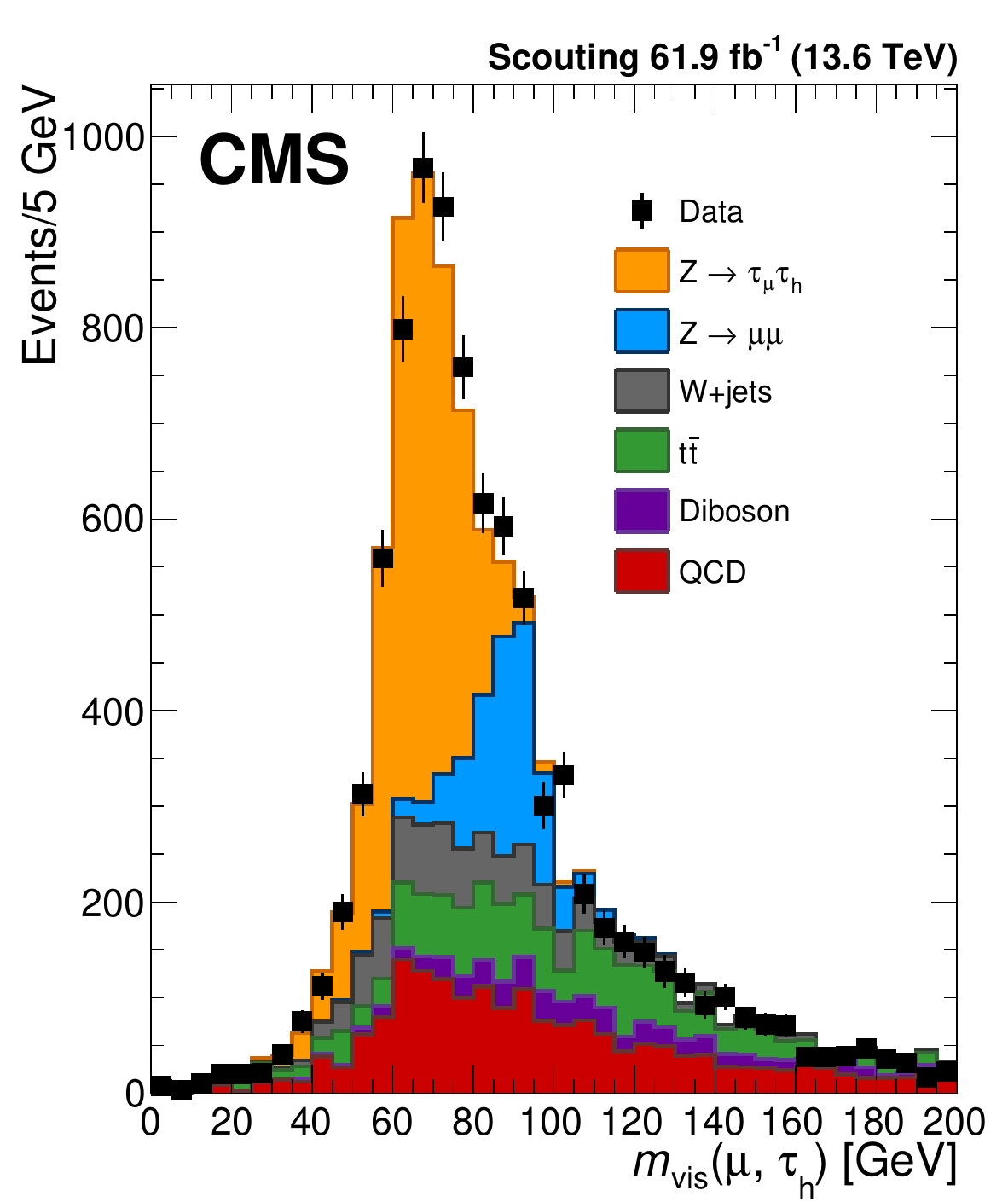}
    \caption{The \mvismutauh distribution in the \PZ boson selection region is shown for data, simulated $\PZ\to\taumu\tauh$ events, and background processes. The QCD background is estimated from same-sign data, while all other backgrounds are taken from simulation. The fitted $\PZ\to\taumu\tauh$ signal agrees with SM expectations.}
    \label{fig:SM_candles}
\end{figure}

Figure~\ref{fig:SM_candles} shows the \mvismutauh distribution for $\PZ\to\taumu\tauh$ candidate events in the \PZ boson observation region. The \PGpz from the \tauh decay primarily fires the ECAL trigger, but extra jets in the events may fire the \HT or jet triggers. The QCD background is estimated from same-sign data after subtracting from it the non-QCD backgrounds taken from simulation. All other background shapes, including $\PZ \to \PGm\PGm$, $\PW$+jets, \ttbar, and dibosons, are estimated using simulation.  The principal resonant background in the \PZ boson observation region, especially prominent in the single prong final state, is $\Zgstar\to\PGm\PGm$+jets with one $\PGm$ misidentified as a single-prong \tauh candidate. The rate of $\Zgstar\to\PGt\PGt$ production, extracted from maximum-likelihood fits to the data distribution, is $2016\pm54\unit{pb}$, consistent with the CMS measurement in the $\PGm\PGm$ channel at $\sqrt{s} = 13.6\TeV$~\cite{CMS:2025sgv}.

\begin{figure*}[htp!]
\centering
\includegraphics[width=0.325\textwidth]{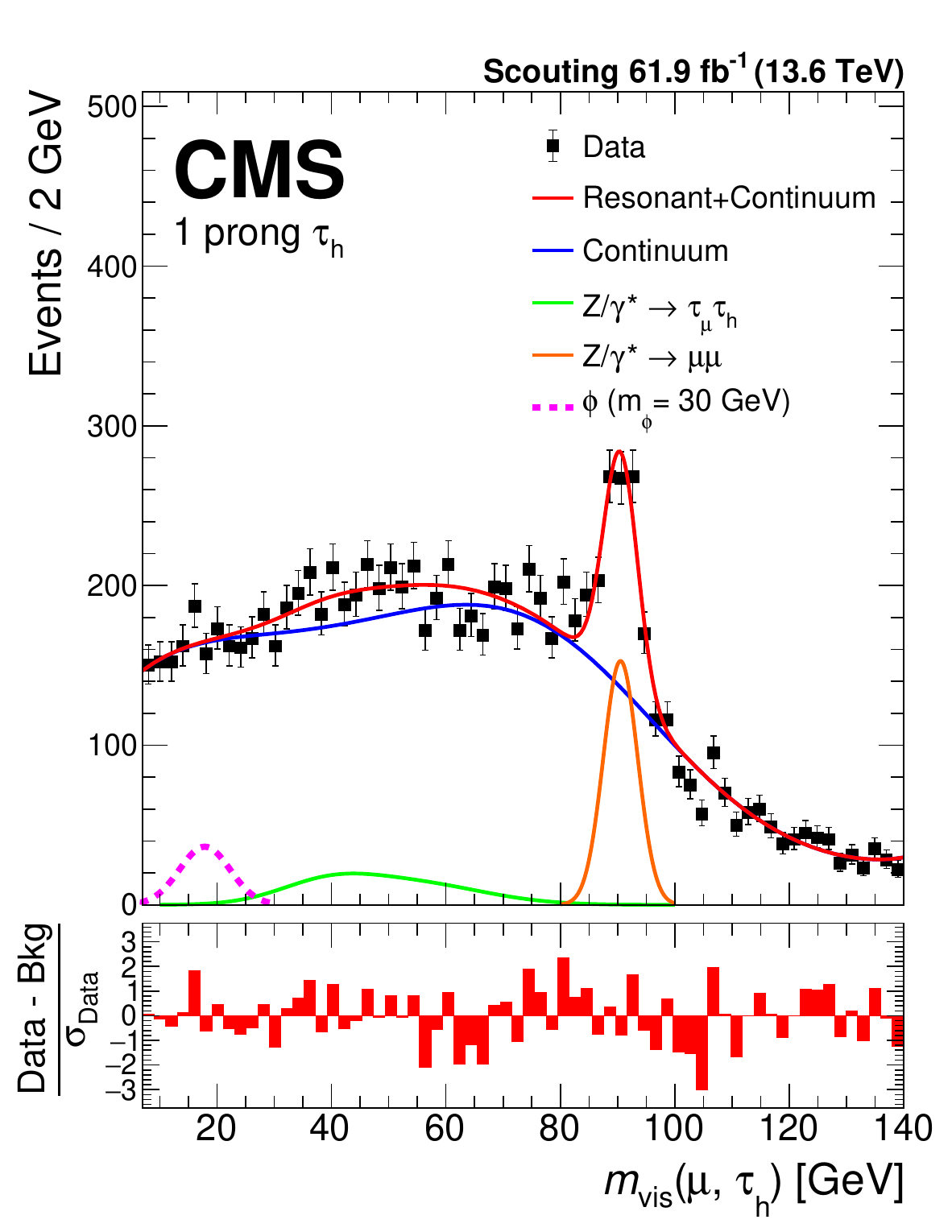}
\includegraphics[width=0.325\textwidth]{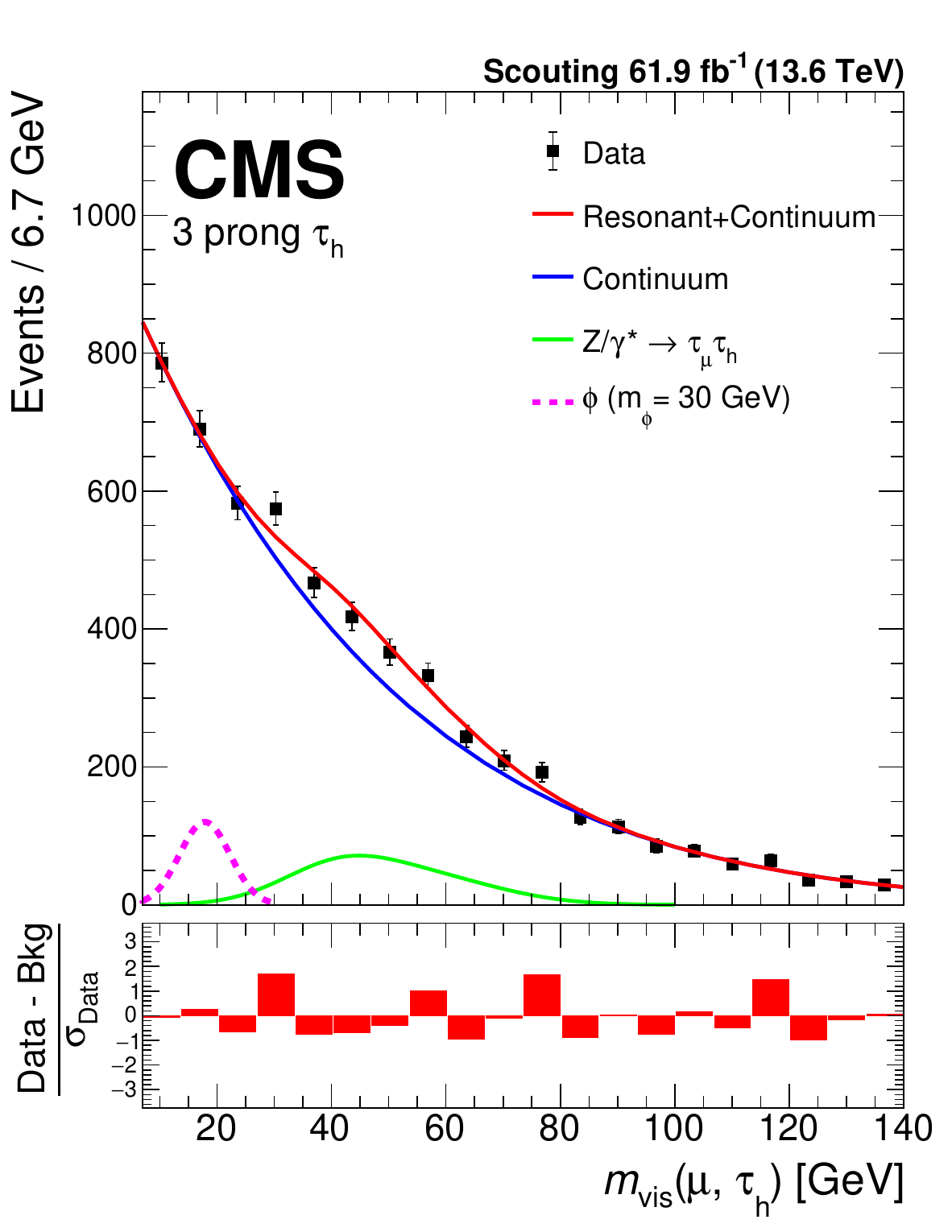}
\includegraphics[width=0.325\textwidth]{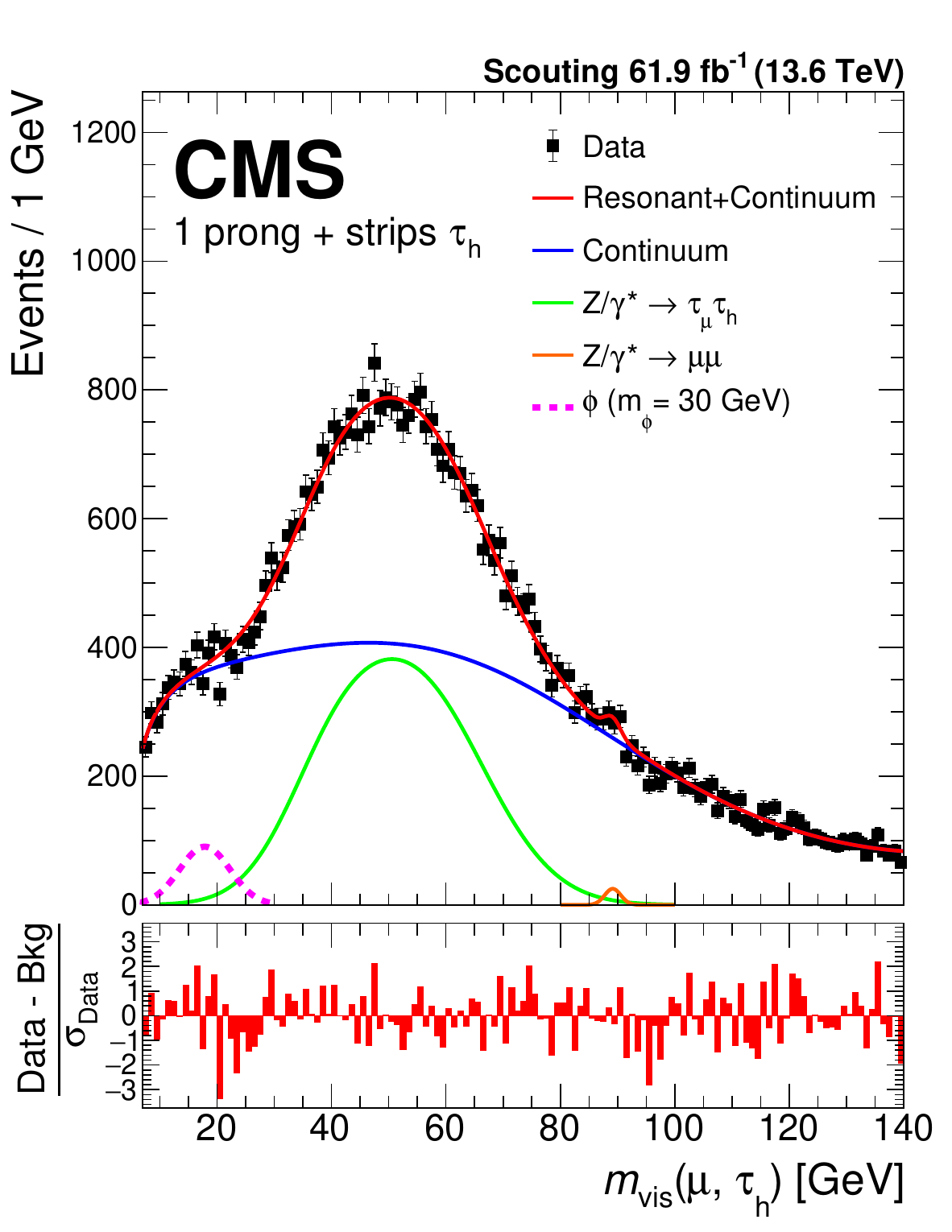}
\caption{
Distributions of \mvismutauh (points with error bars) in the \PGf boson search region, separated by hadronic decay mode with fit results (lines). All plots contain a $\Zgstar\to\PGt\PGt$ component determined by simulations overlaid on the continuum fit. The leftmost plot shows the single-prong final-state with contamination due to $\PZ\to\PGm\PGm$ events (orange) at 91\GeV. The center and right plots show the three-prong and one-prong-plus-one or two-strips respectively. The lower panels of each plot show the differences between the background estimation and the data, divided by the uncertainty in the data points. The \mvismutauh range corresponding to \PGf masses from 20 to 60\GeV is 10 to 40\GeV.}
\label{fig:bkg_fits}
\end{figure*}

In the \PGf search region, we separate the data by \tauh decay mode into one-prong, three-prong, and one-prong-plus-one or two-strip categories and plot their \mvismutauh distributions (Fig.~\ref{fig:bkg_fits}). The fit domain of the \mvis distribution ranges from 6 to 140\GeV, where the upper bound is selected to fully capture the resonant contribution from \Zgstar. The range of \PGf masses (20--60\GeV) corresponds to an \mvismutauh interval from 10 to 40\GeV. The \PGf signal is modeled using a Gaussian fit to the simulated \mvismutauh spectrum. We estimate the resonant and non-resonant (continuum) backgrounds using separate functions. The dominant resonant background is $\Zgstar\to\taumu\tauh+$jets which is modeled by a double-Gaussian function fit to simulation and then convolved with a smearing function. Only the mean and width parameters of the smearing function are allowed to vary in fits to data. Because the $\Zgstar\to\PGm\PGm+$jets falls outside of the search window, we model it with a simple unconstrained Gaussian function. Quarkonium resonances are not considered as backgrounds for $\PGf\to\PGt\PGt$ because they fall below the \mvismutauh range of the search.

We estimate the SM continuum backgrounds with different functions for each \tauh decay group: a fifth-order exponential polynomial ($f(x)=p_{0}\re^{p_{1}x+p_{2}x^{2}+p_{3}x^{3}+p_{4}x^{4}}$) for one-prong, a third-order exponential polynomial ($f(x)=p_{0}\re^{p_{1}x+p_{2}x^{2}}$) for three-prong, and a fifth-order UA2/ATLAS function ($f(x)=p_{0}x^{-p_{1}}e^{p_{2}x+p_{3}x^2+p_{4}x^3}$) for one-prong-plus-strips~\cite{UA2:1990gao,ATLAS:2018qto}. The one-prong and prong-plus-strips categories contain continuum background contributions that have a maximum around 60\GeV in \mvismutauh due to the ECAL trigger threshold. These events originate from single-track electromagnetic contributions coincident with the \tauh candidate, which are not present in the three-prong mode, leading to different continuum background shapes between decay categories. The functional forms and complexities for each category are determined using the Akaike Information Criterion~\cite{Akaike1974}. Figure~\ref{fig:bkg_fits} displays the \mvismutauh distributions from the three \tauh decay groups. Using a continuum plus resonant background-only hypothesis for the fit (red), we extract the $\Zgstar\to\taumu\tauh$ (green) rate, and find that it agrees with the SM expectation within 1 standard deviation of the fit uncertainties in each \tauh decay category. 

To extract a signal yield for each \PGf mass, we perform a maximum-likelihood fit to the data with the continuum and resonant background functions in combination with a Gaussian signal shape consistent with that of the \PGf simulation. This procedure is carried out with the \mvismutauh distributions of all \tauh categories simultaneously. The signal efficiency is of order $10^{-4}$, measured from simulation. Systematic uncertainties in the signal shapes are estimated by examining the average relative differences in the mean and width of the observed $\PZ\to\taumu\tauh$ process compared to simulation shown in Fig.~\ref{fig:SM_candles}. The width and mean uncertainties are 3.7 and 3.5\%, respectively. The $\Zgstar\to\taumu\tauh$ smearing function parameters are constrained such that their effect may only vary to within the shape uncertainties reported. The uncertainties in the \PGf signal mean and width parameters are modeled with Gaussian distributions. All other parameters including those that control the continuum background and $\Zgstar\to\PGm\PGm$ events are unconstrained. We perform goodness-of-fit and bias tests to determine the robustness of the fitting procedure. The bias tests are carried out with pseudodata created from fits to the data using alternate functional forms in the case of a no-signal hypothesis as a template. Signals with varying cross sections were injected to examine any bias in the signal extraction procedure. No significant biases are observed. 

We consider two sources of systematic uncertainty in \tauh reconstruction efficiency in addition to the rate uncertainty in the integrated luminosity. One source is the possible bias on \tauh reconstruction efficiency as a function of \ptvis due to the L1 triggers. To estimate this effect, we compare \tauh collected by the analysis triggers to those collected by an orthogonal L1-dimuon trigger. The variation in efficiency between the triggers was used to estimate this effect. We assign a value of 8\%. Another source of systematic uncertainty is due to differences in \tauh reconstruction between data and simulation. We compare pass/fail ratios of \tauh candidates in simulation with the observed yields in the same pass/fail region of \PZ boson events in data, and assign a value of 10\%. Finally, we assume a 1.4\% uncertainty in the integrated luminosity~\cite{LUM-22-001, CMS-DP-2024-068}. These systematic uncertainties in the rate are treated as nuisance parameters with log-normal priors in the signal extraction procedure.  

\begin{figure}[htpb]
    \centering
    \includegraphics[width=0.4\textwidth]{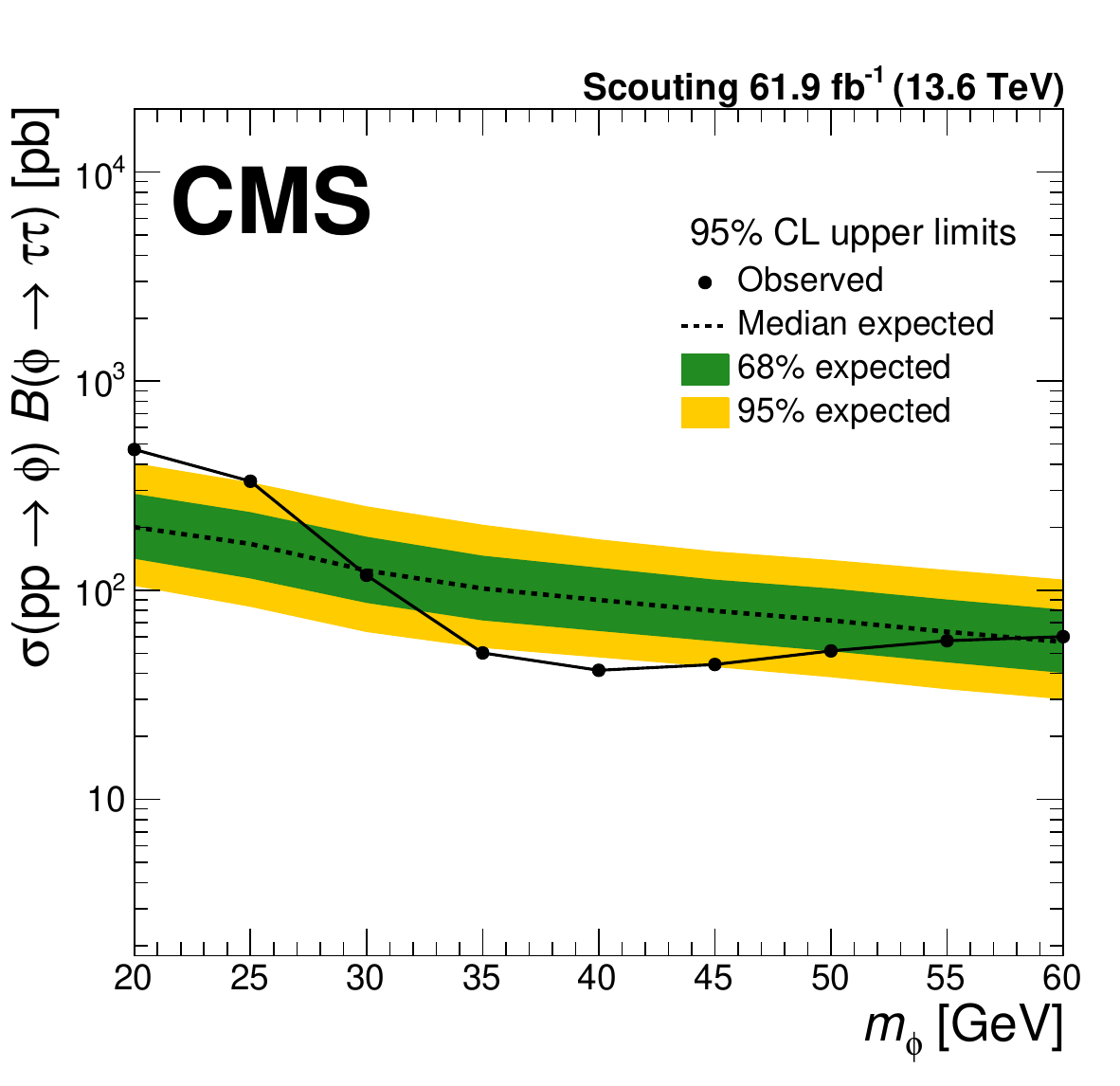}
    \caption{The 95\% \CL upper limits on the product of the cross section and branching fraction, $\sigma(\Pp\Pp\to\PGf)\mathcal{B}(\PGf\to\PGt\PGt)$ in pb. The dotted line represents the median expected limits while the yellow and blue bands indicate the 68 and 95\% expected bands, respectively. The solid line shows the observed limits.}
    \label{fig:limits}
\end{figure}

The fit results are used to set combined 95\% confidence level (\CL) upper limits on the product of the  production cross section and branching fraction $\sigma(\Pp\Pp\to\PGf)\mathcal{B}(\PGf\to\PGt\PGt)$ using the CMS statistical analysis tool \textsc{Combine}~\cite{CMS:2024onh}. We calculate upper limits with a modified frequentist approach based on the \CLs criterion~\cite{CLS1,CLS2} and use an asymptotic approximation for the test statistic~\cite{asymptotic}. The limits are shown as a function of $m_{\PGf}$ in Fig.~\ref{fig:limits}. These results extend the mass region previously analyzed by CMS and ATLAS with a \tauh in the final state from 60 down to 20\GeV, constituting the first ever inclusive search for a spin-zero \PGf with a hadronic \PGt signature produced by gluon fusion in this region. 

In summary, an inclusive search for low-mass spin-zero \PGf bosons decaying to tau leptons with a hadronic signature has been performed in a previously unexplored mass range using proton-proton collision data at $\sqrt{s}=13.6\TeV$, collected with the CMS detector at the LHC in 2022--2023. This search demonstrates the efficacy of a novel low-momentum \tauh reconstruction algorithm implemented in a high-rate data stream. These innovations allow unprecedented sensitivity to new physics with tau leptons. 

\begin{acknowledgments}
We congratulate our colleagues in the CERN accelerator departments for the excellent performance of the LHC and thank the technical and administrative staffs at CERN and at other CMS institutes for their contributions to the success of the CMS effort. In addition, we gratefully acknowledge the computing centers and personnel of the Worldwide LHC Computing Grid and other centers for delivering so effectively the computing infrastructure essential to our analyses. Finally, we acknowledge the enduring support for the construction and operation of the LHC, the CMS detector, and the supporting computing infrastructure provided by the following funding agencies: SC (Armenia), BMBWF and FWF (Austria); FNRS and FWO (Belgium); CNPq, CAPES, FAPERJ, FAPERGS, and FAPESP (Brazil); MES and BNSF (Bulgaria); CERN; CAS, MoST, and NSFC (China); MINCIENCIAS (Colombia); MSES and CSF (Croatia); RIF (Cyprus); SENESCYT (Ecuador); ERC PRG and PSG, TARISTU24-TK10 and MoER TK202 (Estonia); Academy of Finland, MEC, and HIP (Finland); CEA and CNRS/IN2P3 (France); SRNSF (Georgia); BMFTR, DFG, and HGF (Germany); GSRI (Greece); MATE and NKFIH (Hungary); DAE and DST (India); IPM (Iran); SFI (Ireland); INFN (Italy); MSIT and NRF (Republic of Korea); MES (Latvia); LMTLT (Lithuania); MOE and UM (Malaysia); BUAP, CINVESTAV, CONACYT, LNS, SEP, and UASLP-FAI (Mexico); MOS (Montenegro); MBIE (New Zealand); PAEC (Pakistan); MSHE, NSC, and NAWA (Poland); FCT (Portugal); MESTD (Serbia); MICIU/AEI and PCTI (Spain); MOSTR (Sri Lanka); Swiss Funding Agencies (Switzerland); MST (Taipei); MHESI (Thailand); TUBITAK and TENMAK (T\"{u}rkiye); NASU (Ukraine); STFC (United Kingdom); DOE and NSF (USA).
\end{acknowledgments}\section*{Data availability} Release and preservation of data used by the CMS Collaboration as the basis for publications is guided by the  \href{https://doi.org/10.7483/OPENDATA.CMS.1BNU.8V1W}{CMS data preservation, re-use and open access policy}.
\bibliography{auto_generated}
\clearpage
\appendix

\section{Scouting HPS tau reconstruction}\label{App:ScoutngHPSRecoDetails} 

The standard HPS algorithm~\cite{CMS:2018jrd} clusters PF \PGg and \Pepm candidates into ``strips'' in order to identify \PGpz candidates from \tauh decays. We implement the strip clustering algorithm in scouting data using the available $\Pe/\PGg$, because the Scouting HLT algorithm does not separate \Pepm from \PGg. The anti-\kt algorithm~\cite{Cacciari_2008} as implemented in \FASTJET~\cite{Cacciari_2012,Cacciari:2005hq} is utilized to cluster charged hadron (\hpm), \PGpz, \PGmpm, and neutral hadron candidates into jets with a distance parameter equal to 0.4 (AK4 jets). Four \tauh final states are reconstructed: \hpm (one-prong), $\hpm\hmp\hpm$ (three-prong), $\hpm\PGpz$ (one-prong-plus-one-strip), and $\hpm\PGpz\PGpz$ (one-prong-plus-two-strips). All combinations of charged hadron and \PGpz hypotheses that meet the kinematic requirements for each \tauh decay mode are considered in each jet. The \tauh candidates within each jet are ranked by \ptvis and the highest is chosen for that jet. All \tauh candidates at this stage in the event are compared by isolation, as defined in Ref.~\cite{CMS:2018jrd}, and the most isolated candidate is chosen for further analysis. This process selects a single \tauh candidate in each event. 

All constituents of the \tauh candidate must be contained within a region called the \PGt cone. In the standard HPS algorithm, the \PGt cone is constrained to $\Delta R<0.1$ in order to reject QCD backgrounds. However, this limit precludes the reconstruction of many \tauh decays with $\ptvis<18\GeV$. To gain sensitivity to low-\ptvis \tauh decays, the scouting HPS algorithm introduces a dynamic \PGt cone that decreases in size with the increase in \ptvis; this increases sensitivity to low-\ptvis \tauh decays and low-mass $\PGt\PGt$ resonances. This dynamic cone size algorithm also increases the rate of backgrounds misidentified as \tauh candidates. To improve the purity of scouting \tauh candidates, we employ an energy flow neural network based on deep sets~\cite{Komiske:2018cqr} that assigns a likelihood discriminant to identify genuine \tauh decays. This model (\taunet) takes in kinematic and particle identification information about the constituents of the scouting \tauh candidate and the neighboring particles from the AK4 jet in which it is embedded. The variables input to \taunet are as follows: the $\Delta\eta$, $\Delta\phi$, and $\Delta R$ between the AK4 jet axis and its constituents, the natural logarithm of constituent \pt divided by the jet \pt, as well as the natural log of the constituent energy divided by the jet energy. In addition, the particle identification (hadron, $\Pe/\PGg$, or $\PGm$), track impact parameters, and whether or not the constituent has been identified as an element of the \tauh are also inputs to \taunet. We train the model on 1.4 million simulated \tauh decays and an equal number of background events. The backgrounds are selected from scouting data with same-sign $\taumu\tauh$ candidates. The \ptvis spectra of the training data are constructed to be flat in logarithmic space to reduce kinematic bias. 

The highest \pt PF muon in each event is selected as the \taumu candidate. We implement a neural network isolation discriminant (\HLVmu) to select \taumu candidates with low neighboring particle activity. To quantify the isolation of the muon track, we consider a region around it with boundaries determined by $0.001 < \Delta R < 0.4$. The \pt sum of charged hadrons originating from the primary vertex (PV), charged hadrons originating away from the PV, neutral hadrons, and $\Pe/\PGg$ are scaled by the \ptvis of the \PGm candidate and input to the neural network along with the transverse impact parameter $d_\mathrm{xy}$, longitudinal impact parameter $d_\mathrm{z}$, and PV assignment of the scouting PF \PGm candidate. Constituents of the \tauh candidate are not considered in the isolation calculation of the \PGm, and vice-versa. This allows the \tauh and \PGm to exist in close proximity, allowing sensitivity to highly Lorentz-boosted and merged $\taumu\tauh$ systems.

\cleardoublepage \section{The CMS Collaboration \label{app:collab}}\begin{sloppypar}\hyphenpenalty=5000\widowpenalty=500\clubpenalty=5000\cmsinstitute{Yerevan Physics Institute, Yerevan, Armenia}
{\tolerance=6000
A.~Hayrapetyan, V.~Makarenko\cmsorcid{0000-0002-8406-8605}, A.~Tumasyan\cmsAuthorMark{1}\cmsorcid{0009-0000-0684-6742}
\par}
\cmsinstitute{Institut f\"{u}r Hochenergiephysik, Vienna, Austria}
{\tolerance=6000
W.~Adam\cmsorcid{0000-0001-9099-4341}, L.~Benato\cmsorcid{0000-0001-5135-7489}, T.~Bergauer\cmsorcid{0000-0002-5786-0293}, M.~Dragicevic\cmsorcid{0000-0003-1967-6783}, P.S.~Hussain\cmsorcid{0000-0002-4825-5278}, M.~Jeitler\cmsAuthorMark{2}\cmsorcid{0000-0002-5141-9560}, N.~Krammer\cmsorcid{0000-0002-0548-0985}, A.~Li\cmsorcid{0000-0002-4547-116X}, D.~Liko\cmsorcid{0000-0002-3380-473X}, M.~Matthewman, J.~Schieck\cmsAuthorMark{2}\cmsorcid{0000-0002-1058-8093}, R.~Sch\"{o}fbeck\cmsAuthorMark{2}\cmsorcid{0000-0002-2332-8784}, M.~Shooshtari\cmsorcid{0009-0004-8882-4887}, M.~Sonawane\cmsorcid{0000-0003-0510-7010}, W.~Waltenberger\cmsorcid{0000-0002-6215-7228}, C.-E.~Wulz\cmsAuthorMark{2}\cmsorcid{0000-0001-9226-5812}
% ...
\par}
\vskip\cmsinstskip
\dag:~Deceased\\
$^{1}$Also at Yerevan State University, Yerevan, Armenia\\
$^{2}$Also at TU Wien, Vienna, Austria\\
% ...
\end{sloppypar}
%%% END EDITABLE REGION %%%
% skeleton_end
\end{document}